\documentclass[twocolumn,showpacs,preprintnumbers]{revtex4}

\usepackage{amsmath,amssymb,graphicx,cancel}
\usepackage[dvips]{color}

\begin{document}
\title{A powerful tool for measuring
Higgs boson associated Lepton Flavour Violation}
\preprint{IC/2009/002, UT-HET 021}
\author{Shinya Kanemura}
\email{kanemu@sci.u-toyama.ac.jp}
\affiliation{Department of Physics, University of Toyama,
3190 Gofuku, Toyama 930-8555, Japan}
\author{Koji Tsumura}
\email{ktsumura@ictp.it}
\affiliation{The Abdus Salam ICTP of
UNESCO and IAEA, Strada Costiera 11, 34014 Trieste, Italy}
\begin{abstract}
In models with extended Higgs sectors, Higgs-boson-mediated
Lepton Flavour Violation (LFV) can naturally appear.
We study the physics potential of an electron-photon collider
on searching LFV processes
$e^-\gamma\to\ell^-\varphi\;(\ell=\mu,\tau; \varphi=H, A)$
where $H$ and $A$ are extra CP even and odd Higgs bosons, respectively,
in the minimal supersymmetric standard model and the effective
two Higgs doublet model.
The production cross section can be significantly large for
the maximal allowed values of the LFV coupling constants under the
current experimental data.
Present experimental upper bounds on the effective LFV coupling constants
would be considerably improved by searching these processes, which
would be better than MEG and COMET experiments and also those at LHCb
and SuperKEKB.
Moreover, one can separately measure chirality of effective LFV coupling
constants via these processes by selecting electron polarizations.
\end{abstract}
\maketitle

Lepton Flavour Violation (LFV) for charged leptons is, if it exists,
a clear signature of new physics beyond the Standard Model (SM).
Various new physics scenarios such as supersymmetric extensions of the
SM can naturally predict observable phenomena of LFV,
whose details would relate to
fundamental flavour structures at very high energies.
Hence, experimental detection of LFV phenomena cannot only provide an
evidence of new physics, but also its detailed information can be used
to distinguish new physics models.
Phenomenologically there are two kinds of LFV processes;
i.e., the gauge boson $(\gamma,Z)$ mediation and the Higgs boson mediation.

LFV coupling constants in association with Higgs bosons
can appear in models with extended Higgs sectors.
In supersymmetric extensions of the SM whose Higgs sectors have
at least two scalar isospin doublet fields, LFV Yukawa interactions
can be radiatively induced from slepton mixing~\cite{Ref:YnuRGE,Ref:slmix}.
Flavour mixing between ``left-handed'' sleptons may be induced
by the quantum effect via the neutrino Yukawa coupling constants in the
minimal supersymmetric SM with heavy right-handed neutrinos (MSSMRN)
when flavour blind structure is assumed at the grand unification scale~\cite{Ref:YnuRGE}.
In a general framework of supersymmetric SMs,
not only mixing between``left-handed'' sleptons but also
that between ``right-handed''sleptons can be considered~\cite{Ref:slmix}.
Such a difference in patterns of LFV in various models
can be studied by measuring effective LFV parameters in the Yukawa
interaction as well as those in the effective LFV gauge interactions.

The effective LFV parameters have been tested
at the experiments for rare decay processes of muons and tau leptons, 
and their upper bounds have
been obtained~\cite{Ref:BKDER,Ref:Sher,Ref:KOT}.
They are expected to be improved at
PSI MEG~\cite{Ref:MEG} and J-PARC COMET~\cite{Ref:COMET}
experiments via rare decays of muons, and at CERN LHCb~\cite{Ref:LHCb}
and KEK super-B factory~\cite{Ref:SuperB} via rare decays of tau leptons.
In addition, high energy collider experiments
at the CERN Large Hadron Collider
(LHC)~\cite{Ref:LHC}, the International Linear Collider
(ILC)~\cite{Ref:ILC} and the Neutrino Factory~\cite{Ref:nuFact}
can also be useful to search
Higgs-boson-mediated $\mu$--$\tau$ and $e$--$\tau$ mixing~\cite{Ref:LFVhdecayLHC,Ref:KMOSTT,Ref:KOT, Ref:KKKO}.

In this letter, we discuss the possibility that
a high energy $e\gamma$ collider, which can be realized as an optional
experiment at the ILC, can be a powerful tool for measuring
Higgs-boson-mediated LFV parameters in two Higgs doublet models (THDMs) including
Minimal Supersymmetric SMs (MSSMs).
We consider the processes
$e^- \gamma\to \ell^- \varphi\; (\ell=\mu, \tau; \varphi=h, H, A)$,
which contain the LFV couplings of $e^-\ell^+\varphi$,
where $h$, $H$ and $A$ are neutral Higgs bosons.
Advantages of these processes turn out to be the following.
i) The total cross sections for $e^-\gamma \to \tau^-A$
($e^-\gamma \to \mu^-A$) can be about $10.6$ fb ($7.3$ fb) for the collision 
energy of $e^-e^-$ system to be $\sqrt{s_{ee}}=522$ GeV ($471$ GeV), 
where $m_A^{}=350$ GeV, $\tan\beta=50$ and allowed maximal values 
of the LFV parameters under the constraint from the current experimental 
data are taken, where $\tan\beta$ is the ratio of vacuum expectation values 
for the two Higgs doublets.
ii) These processes are basically background free.
Therefore, the current upper bound on the LFV parameters
in the effective Yukawa interaction can be improved by several orders
of magnitude by assuming moderate properties for $e\gamma$ colliders.
iii) Information of the Higgs-boson-mediated LFV couplings can directly
be extracted. Produced Higgs bosons can be reconstructed by tagging its
decay product $b\overline{b}$.
iv) The use of polarized beams for incident electrons
makes it possible to discriminate the chirality of the LFV parameters,
thereby we may be able to distinguish scenarios for more fundamental physics models.

The effective Yukawa interaction for charged leptons
is given in the general framework of the THDM by~\cite{Ref:Sher,Ref:KOT}
\begin{align}
\mathcal{L}_{\text{lepton}}
=& -\overline{{\ell_R}_i}\left\{Y_{\ell_i}\delta_{ij}\Phi_1
+\left(Y_{\ell_i}\epsilon_{ij}^L+\epsilon_{ij}^RY_{\ell_j}\right)
\Phi_2\right\}\cdot L_j\nonumber\\
&+\text{H.c.},
\label{Eq:yukawa}
\end{align}
where ${\ell_R}_i (i=1$--$3)$ represent isospin singlet fields of
right-handed charged leptons, $L_i$ are isospin doublets of
left-handed leptons, $Y_{\ell_i}$ are the Yukawa coupling constants
of $\ell_i$, and $\Phi_1$ and $\Phi_2$ are the scalar iso-doublets
with hypercharge $Y=1/2$.
Parameters $\epsilon_{ij}^X (X=L,R)$ can induce LFV interactions
in the charged lepton sector in the basis of the mass eigenstates.
In Model II THDM~\cite{Ref:HHG}, $\epsilon_{ij}^X$ vanishes at the
tree level, but it can be generated radiatively by new physics
effects~\cite{Ref:slmix}.

The effective Lagrangian can be rewritten in terms of physical Higgs
boson fields. There are eight degrees of freedom in two scalar
doublets $\Phi_1$ and $\Phi_2$. Three of them are absorbed
by massive gauge bosons via the Higgs mechanism, and remaining
five are physical Higgs bosons. Assuming the CP invariant Higgs
sector, there are two CP even Higgs bosons $h$ and $H$
$(m_h^{}<m_H^{})$, one CP odd state $A$ and a pair of charged
Higgs bosons $H^\pm$.
From Eq.~\eqref{Eq:yukawa}, interaction terms can be deduced to~\cite{Ref:slmix,Ref:KOT}
\begin{align}
&\mathcal{L}_{e\text{LFV}}
= -\frac{m_{\ell_i}}{v\cos^2\beta}
\left(\kappa^L_{i1}\overline{\ell_i}\text{P}_L^{}e
+\kappa^R_{1i}\overline{e}\text{P}_L^{}\ell_i\right)\nonumber\\
&\times\left\{\cos(\alpha-\beta)h+\sin(\alpha-\beta)H-i\,A\right\}
+\text{H.c.},
\end{align}
where $\text{P}_L^{}$ is the projection operator to the
left-handed fermions, $m_{\ell_i}$ are mass eigenvalues of charged leptons,
$v=\sqrt{2} \sqrt{\langle\Phi_1^0\rangle^2+\langle\Phi_2^0\rangle^2}$
($\simeq 246$ GeV),
$\alpha$ is the mixing angle between the CP even Higgs bosons, and
$\tan\beta\equiv\langle\Phi_2^0\rangle/\langle\Phi_1^0\rangle$.
The LFV parameters $\kappa_{ij}^{X} (X=L,R)$ relate to the original
parameters as~\cite{Ref:YnuRGE}
\begin{align}
\kappa_{ij}^X=-\frac{\epsilon_{ij}^X}{\left\{1+
\left(\epsilon_{33}^L+\epsilon_{33}^R\right)\tan\beta\right\}^2},
\end{align}
for $\epsilon_{ij}^X\tan\beta\ll 1$.

Once a new physics model is assumed, $\kappa_{ij}^X$ can be predicted
as a function of the model parameters. In supersymmetric SMs,
LFV Yukawa coupling constants can be radiatively generated
by slepton mixing.
Magnitudes of the LFV parameters
$\kappa^{X}_{ij}$ can be calculated as a function of the parameters
of the slepton sector. For the scale of the dimensionful parameters
in the slepton sector to be of TeV scales, we typically obtain
$|\kappa^{X}_{ij}|^2 \sim (1$--$10) \times 10^{-7}$~\cite{Ref:YnuRGE,Ref:slmix}.
In the MSSMRN only $\kappa^L_{ij}$ are generated by the
quantum effect via the neutrino Yukawa couplings assuming
flavour conservation at the scale of right-handed neutrinos.

Current experimental bounds on the effective LFV parameters
$\kappa_{ij}^X$ are obtained from the data of
non-observation for various LFV processes~\cite{Ref:Rare}.
For $e$--$\tau$ mixing, we obtain the upper bound
from the semi-leptonic decay $\tau\to e\eta$~\cite{Ref:Sher};
\begin{align}\hspace{-1ex}
|\kappa^L_{31}|^2+|\kappa^R_{13}|^2&\lesssim 6.4\times 10^{-6}
\left(\frac{50}{\tan\beta}\right)^6
\left(\frac{m_A^{}}{350\text{GeV}}\right)^4,\label{eq:kapmax13}
\end{align}
for $\tan\beta \gtrsim 20$ and $m_A^{} \simeq m_H^{} \gtrsim 160$ GeV (with
$\sin(\beta-\alpha)\simeq 1$).
The most stringent bound on $e$--$\mu$ mixing is derived from
$\mu\to e\gamma$ data~\cite{Ref:PDG} as
\begin{align}
\frac49|\kappa^L_{21}|^2+|\kappa^R_{12}|^2&\lesssim 4.3\times 10^{-4}
\left(\frac{50}{\tan\beta}\right)^6
\left(\frac{m_A^{}}{350\text{GeV}}\right)^4, \label{eq:kapmax12}
\end{align}
for $\tan\beta \gtrsim 20$ and $m_A^{} \simeq m_H^{} \gtrsim 160$ GeV (with
$\sin(\beta-\alpha)\simeq 1$).
\begin{figure}[tb]
\includegraphics[width=7.5cm]{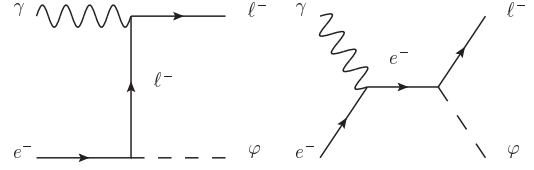}
\caption{The Feynman diagrams for
$e^-\gamma \to \ell^- \varphi\;(\ell=\mu, \tau;
\varphi=H,A)$.}
\label{FIG:Feyn}
\end{figure}
The upper bound on $(4/9)|\kappa_{21}^L|^2+|\kappa_{12}^R|^2$ is expected to
be improved at future experiments such as MEG and COMET for rare muon decays
by a factor of $10^{2\text{--}3}$, while that on
$|\kappa_{31}^L|^2+|\kappa_{13}^R|^2$ is by $10^{1\text{--}2}$
at LHCb and SuperKEKB via rare tau
decays~\cite{Ref:MEG,Ref:COMET,Ref:LHCb,Ref:SuperB}.

\begin{figure*}[tb]
\includegraphics[width=7.5cm]{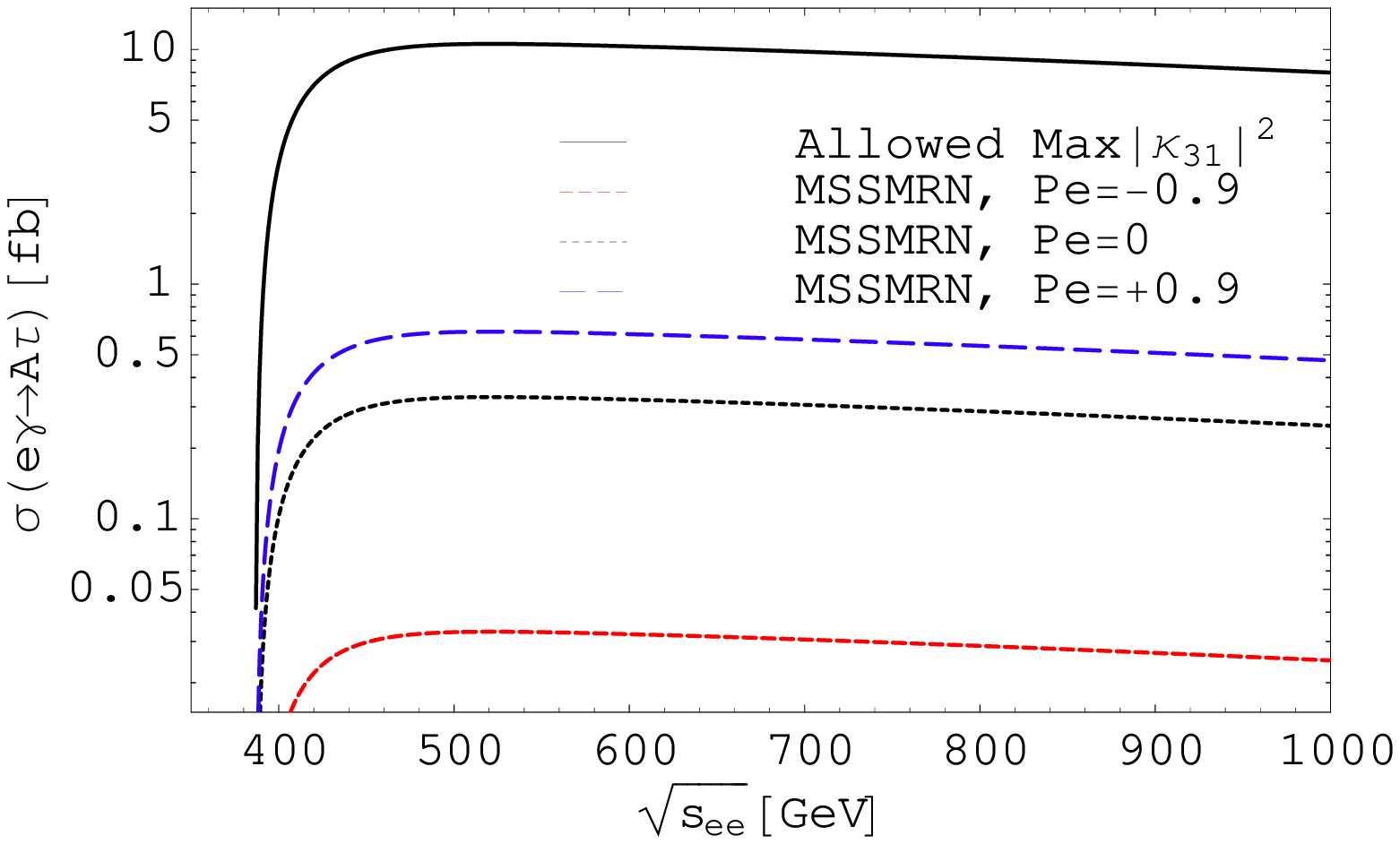}
\includegraphics[width=7.5cm]{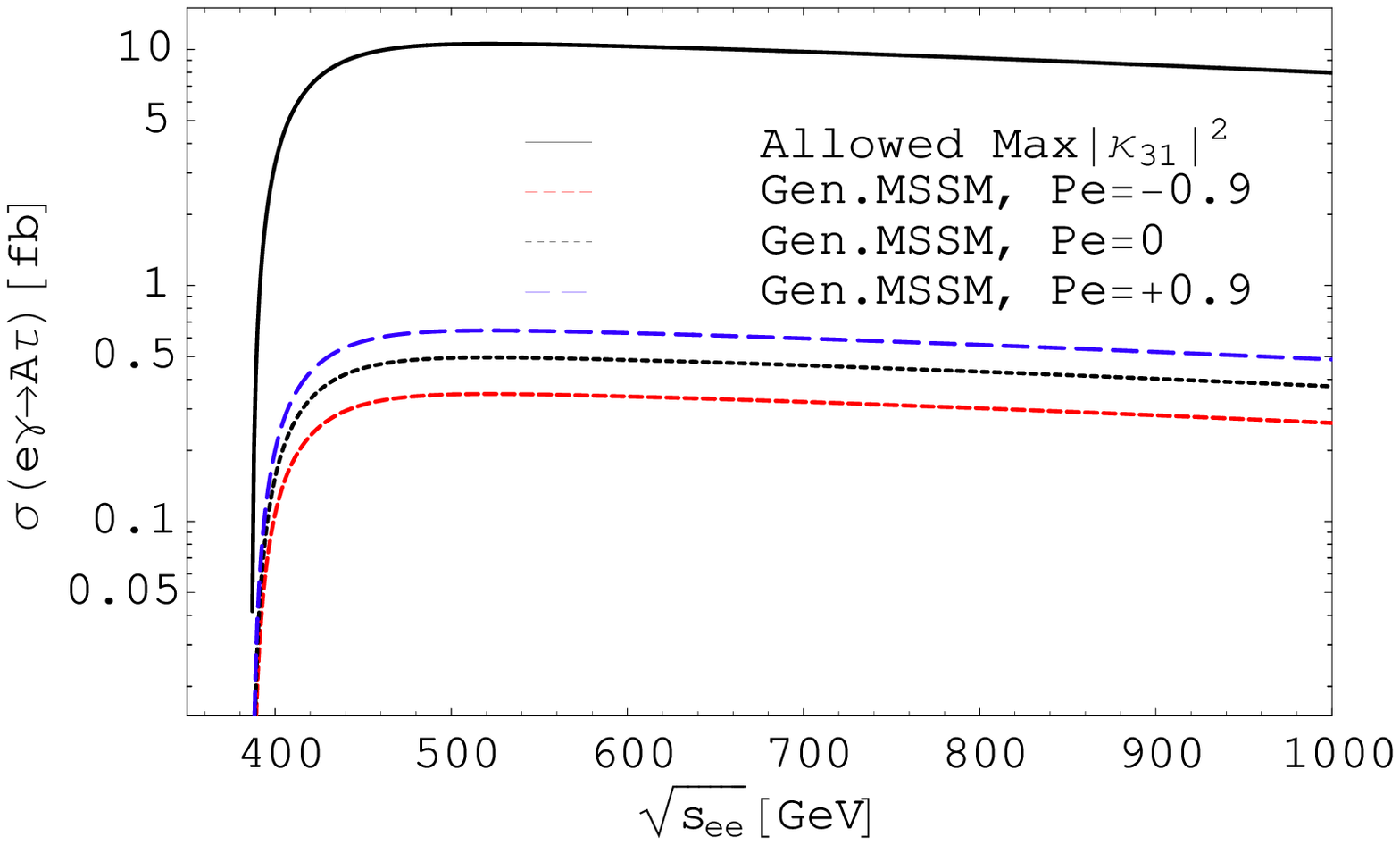}
\caption{
 The production cross section of $e^-\gamma \to \tau^- A$ as a function of
the center-of-mass energy  $\sqrt{s_{ee}}$ of the electron-electron system.
Final state leptons can be detected in the range $\epsilon\le\theta\le\pi-\epsilon$
where $\epsilon=20$ mrad.
Solid curve represents the result in the THDM with the maximal allowed value of
$|\kappa_{31}|^2$ under the current experimental data in Eq.~\eqref{eq:kapmax13}
in both figures.
In the left figure, the result with a set of the typical values of
$|\kappa^L_{31}|^2$ and $|\kappa^R_{13}|^2$
in the MSSMRN is shown [$(|\kappa^L_{31}|^2, |\kappa^R_{13}|^2)=(2\times 10^{-7}, 0)$] for the polarization of  the incident electron beam to be
 $P_e=-0.9$ (dashed), $+0.9$ (long dashed), and $P_e=0$ (dotted).
In the right figure, the result with a set of
$(|\kappa^L_{31}|^2, |\kappa^R_{13}|^2)=(2\times 10^{-7}, 1\times
 10^{-7})$ is shown in the general framework of the MSSM for
 $P_e=-0.9$ (dashed), $P_e=+0.9$ (long dashed), and $P_e=0$ (dotted).
 }
\label{FIG:EGamAtau}
\end{figure*}

We now discuss the lepton flavour violating Higgs boson production
processes $e^-\gamma \to \ell^- \varphi\;(\ell=\mu, \tau; \varphi=h, H, A)$
in $e\gamma$ collisions.
The Feynman diagrams for the sub processes are depicted in FIG.~\ref{FIG:Feyn}.
The differential cross section is calculated by using the
effective LFV parameters $\kappa_{ij}^X$  as
\begin{align}
%
&\frac{d\widehat{\sigma}_{e^-\gamma\to\ell^-_i\varphi}(\sqrt{s_{e\gamma}})}{d\cos\theta}\nonumber\\
&=\frac{G_F\alpha_\text{EM}^{}m_\ell^2\beta_{\ell\varphi}}{16\sqrt2s_{e\gamma}}
\frac{\left|\kappa_{i1}\right|^2}{\cos^4\beta}
\frac{\eta_-(\eta_+^2+4z^2)-16z\,m_\ell^2/s_{e\gamma}}{\eta_-^2},
\end{align}
where we have neglected the mass of electrons,
$z=(m_{\ell_i}^2-m_\varphi^2)/s_{e\gamma}$,
$\beta_{\ell\varphi}=\sqrt{\lambda(m_{\ell_i}^2/s_{e\gamma},
m_\varphi^2/s_{e\gamma})}$
with $\lambda(a,b)=1+a^2+b^2-2a-2b-2a b$.
The functions are defined as $\eta_\pm=1+z\pm\beta_{\ell\varphi}\cos\theta$
where $\theta$ is the scattering angle of the outgoing lepton from
the beam direction.
The effective LFV parameters $|\kappa_{i1}|^2$ can be written by
\begin{align}
|\kappa_{i1}|^2 = &\left[  |\kappa^L_{i1}|^2 (1-P_e)
+ |\kappa^R_{1i}|^2 (1+P_e)\right] \nonumber \\
&\times
\begin{cases}
\cos^2(\alpha-\beta)&\text{ for }h\\
\sin^2(\alpha-\beta)&\text{ for }H\\
1&\text{ for }A\end{cases},
\end{align}
where $P_e$ is the polarization of the incident electron beam:
$P_e=-1$ ($+1$) represents that electrons in the beam are
$100\%$ left- (right-) handed.

At the ILC, a high energy photon beam can be obtained by Compton
backward-scattering of laser and an electron beam~\cite{Ref:PLC}.
The full cross section can be evaluated from that for the sub process by convoluting
with the photon structure function as~\cite{Ref:PLC}
\begin{align}
\sigma\left(\sqrt{s_{ee}}\right)
=\int_{x_{min}}^{x_{max}}dx\,F_{\gamma/e}(x)\,
\widehat{\sigma}_{e^-\gamma\to\ell^-\varphi}
(\sqrt{s_{e\gamma}}),
\end{align}
where $x_{max}=\xi/(1+\xi)$, $x_{min}=(m_\ell^2+m_\varphi^2)/s_{ee}$,
$\xi=4E_e\omega_0/m_e^2$ with $\omega_0$ to be the frequency of the
laser and $E_e$ being the energy of incident electrons,
and $x=\omega/E_e$ with $\omega$ to be the photon energy in the scattered photon beam.
The photon distribution function is given by~\cite{Ref:PLC}
\begin{align}
&F_{\gamma/e}(x)= \frac1{D(\xi)}\nonumber\\
&\times \left\{1-x+\frac1{1-x}-\frac{4x}{\xi(1-x)}
+\frac{4x^2}{\xi^2(1-x)^2}\right\},
\end{align}
with
\begin{align}
D(\xi)=&\left(1-\frac4{\xi}-\frac8{\xi^2}\right)\ln(1+\xi)
+\frac12+\frac8{\xi}-\frac1{2(1+\xi)^2}.
\end{align}
We note that when $\sin(\beta-\alpha)\simeq 1$ and $m_H^{} \simeq m_A^{}$
(In the MSSM, this automatically realizes for $m_A \gtrsim 160$ GeV)
signal from both $e^-\gamma\to\ell^-H$
and $e^-\gamma\to\ell^-A$ can be used to measure the LFV
parameters, while the cross section for $e^-\gamma\to\ell^-h$ is
suppressed.

In FIG.~\ref{FIG:EGamAtau}, we show the full cross sections of
$e^-\gamma \to \tau^- A$ as a function of the center-of-mass energy
of the $e^-e^-$ system for $\tan\beta=50$ and $m_A^{}=350$ GeV.
Scattered leptons mainly go into the forward direction, however
most of events can be detected by imposing the escape cut
$\epsilon\le\theta\le\pi-\epsilon$
where $\epsilon=20$ mrad~\cite{Ref:Anipko}.
The cross section can be around $10$ fb with the maximal allowed values
for $|\kappa_{31}|^2$ under the constraint from the $\tau\to e\eta$ data
in Eq.~(\ref{eq:kapmax13}).
The results correspond that, assuming the integrated luminosity of the $e\gamma$
collision to be $500$ fb$^{-1}$ and the tagging efficiencies of a $b$
quark and a tau lepton
to be $60\%$ and $30\%$, respectively,
about $10^3$ of $ \tau^- b\bar b$ events can be observed as the signal,
where we multiply factor of two by adding both
$e^-\gamma\to\ell^-A\to\ell^-b\overline{b}$ and
$e^-\gamma\to\ell^-H\to\ell^-b\overline{b}$.
Therefore, we can naively say that non-observation of the signal
improves the upper bound for the $e$-$\tau$ mixing by $2$--$3$ orders of
magnitude if the backgrounds are suppressed.
In FIG.~\ref{FIG:EGamAtau} (left), those with a set of the typical values of
$|\kappa^L_{31}|^2$ and $|\kappa^R_{13}|^2$ in the MSSMRN are shown for
 $P_e=-0.9$ (dashed), $P_e=+0.9$ (long dashed), and $P_e=0$ (dotted), where we take
 $(|\kappa^L_{31}|^2, |\kappa^R_{13}|^2)=(2\times 10^{-7}, 0)$.
The cross sections are sensitive to the polarization of
the electron beam. They can be as large as $0.5$ fb for $P_e=-0.9$,
while it is around $0.03$ fb for $P_e=+0.9$.
In FIG.~\ref{FIG:EGamAtau} (right), the results with
$(|\kappa^L_{31}|^2, |\kappa^R_{13}|^2)=(2\times 10^{-7}, 1\times
 10^{-7})$ in general supersymmetric models
 are shown for each polarization of the incident electrons.
The cross sections are a few times $1$ fb and not sensitive for polarizations.
Therefore, by using the polarized beam of the electrons
we can separately measure $|\kappa_{31}^L|^2$ and $|\kappa_{13}^R|^2$ and
 distinguish fundamental models with LFV.

\begin{figure*}[tb]
\includegraphics[width=8cm]{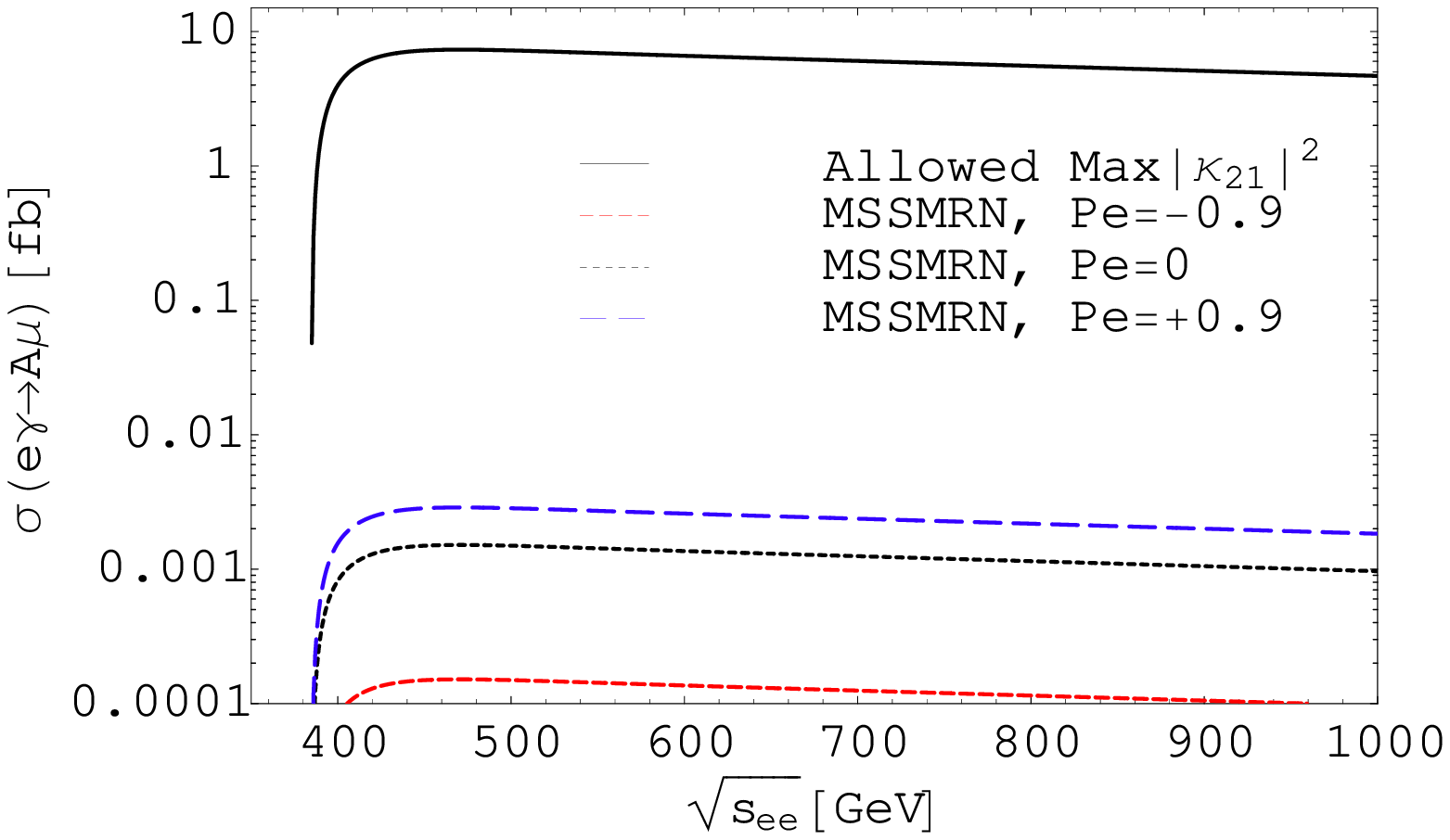}
\includegraphics[width=8cm]{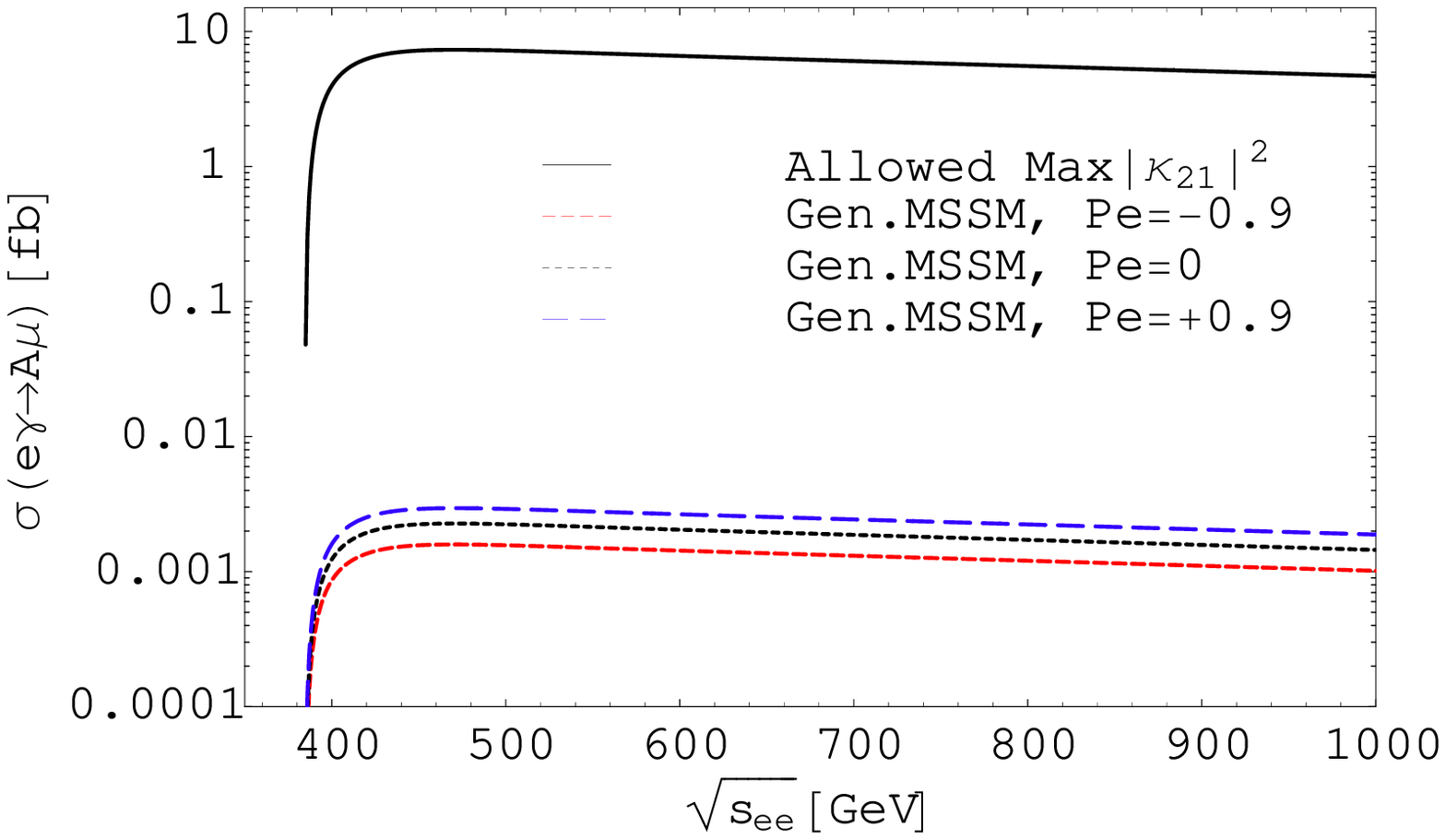}
\caption{
 The production cross section of $e^-\gamma \to \mu^- A$ as a function of
the center-of-mass energy  $\sqrt{s_{ee}}$ of the electron-electron system.
Scattering angle of leptons are restricted as $\epsilon\le\theta\le\pi-\epsilon$
where $\epsilon=20$ mrad.
Solid curve represents the result in the THDM with the maximal allowed value of
$|\kappa_{21}|^2$ under the current experimental data in Eq.~\eqref{eq:kapmax12}
in both figures.
In the left figure, the result with a set of the typical values of
 $|\kappa^L_{21}|^2$ and $|\kappa^R_{12}|^2$
in the MSSMRN is shown [$(|\kappa^L_{21}|^2, |\kappa^R_{12}|^2)=(2\times 10^{-7}, 0)$] for
 the polarization of  the incident electron beam to be
 $P_e=-0.9$ (dashed), $+0.9$ (long dashed), and $P_e=0$ (dotted).
In the right figure, the result with
$(|\kappa^L_{21}|^2, |\kappa^R_{12}|^2)=(2\times 10^{-7}, 1\times
 10^{-7})$
 is shown in the general framework of the MSSM for
 $P_e=-0.9$ (dashed), $P_e=+0.9$ (long dashed), and $P_e=0$ (dotted).
}
\label{FIG:EGamAmu}
\end{figure*}
In FIG.~\ref{FIG:EGamAmu}, the full cross sections of $e^-\gamma \to
\mu^- A$ are shown for $\tan\beta=50$ and $m_A^{}=350$ GeV.
Those with the maximally allowed values for
$|\kappa_{21}|^2=|\kappa^L_{21}|^2+|\kappa^R_{12}|^2$
from the $\mu \to  e \gamma$ data in Eq.~(\ref{eq:kapmax12}) can be $7.3$ fb
where we here adopted the same escape cut as before discussed~
\footnote{If $10$ mrad for the cut is taken instead of $20$ mrad, the
numbers of events are slightly enhanced; $10.6$ fb to $11.0$ fb
($7.3$ fb to $8$ fb) for the $\tau$-$\varphi$  ($\mu$-$\varphi$) process.}.
This means that about a few times $10^3$ of the signal $\mu^- b\bar b$ can be
produced for the integrated luminosity of the $e\gamma$ collision to be
$500$ fb$^{-1}$, assuming tagging efficiencies to be $60\%$ for a $b$
quark and $100\%$ for a muon, and using both $e^-\gamma \to
\mu^- A$ and $e^-\gamma \to \mu^- H$.
These results imply that $e\gamma$ collider can improve the bound on
the $e$-$\mu$ by a factor of $10^{2-3}$.
Obtained sensitivity can be as large as those at
undergoing MEG and projected COMET experiments.
Because of the different dependencies on the parameters in the model,
$\mu\to e\gamma$ can be sensitive than the LFV Higgs boson production
for very high $\tan\beta (\gtrsim 50)$ with fixed Higgs boson mass. 
We also note that rare decay processes can measure the effect of other 
LFV origin when Higgs bosons are heavy. Therefore, both the direct and 
the indirect measurements of LFV processes are complementary to each other.
In FIG.~\ref{FIG:EGamAmu} (left), those in the MSSMRN are shown for
 $P_e=-0.9$ (dashed), $P_e=+0.9$ (long dashed), and $P_e=0$ (dotted), where we take
 $(|\kappa^L_{21}|^2, |\kappa^R_{12}|^2)=(2\times 10^{-7}, 0)$.
They can be as large as a few times $10^{-3}$ fb for $P_e=-0.9$ and
$P_e=0$, while it is around $10^{-4}$ fb for $P_e=+0.9$.
In FIG.~\ref{FIG:EGamAmu} (right), the results with
$(|\kappa^L_{21}|^2, |\kappa^R_{12}|^2)=(2\times 10^{-7}, 1\times
10^{-7})$ are shown in general supersymmetric models in a similar
manner.

It is understood that these processes are clear against backgrounds.
For the processes of  $e^-\gamma \to \tau^- \varphi \to \tau^- b\bar b$.
The tau lepton decays into various hadronic and
leptonic modes.
The main background comes from $e^-\gamma\to W^-Z\nu$, whose cross
section is of the order of $10^2$ fb.
The backgrounds can strongly be suppressed
by the invariant mass cut for $b\bar b$.
The backgrounds for the
process $e^-\gamma \to \mu^- \varphi \to \mu^- b\bar b$ also comes from
$e^-\gamma\to W^-Z\nu\to \mu^-b\overline{b}\nu\overline{\nu}$ which
is small enough. Signal to background ratios are better than
${\mathcal O}(1)$ before kinematic cuts. They are easily
improved by the invariant mass cut, so that our signals can be
almost background free.
A detailed simulation study will clarify this observation.

We have discussed the processes
$e^-\gamma\to\ell^-\varphi\;(\ell=\mu,\tau; \varphi=H, A)$.
Many new physics models can violate lepton flavour in the scalar sector.
Cross sections for the lepton flavour violating Higgs boson production
production at the $e\gamma$ collider can be substantial for the maximal
allowed values  under the current experimental data.
Measurements of these processes can significantly improve the present upper
bounds of the LFV Yukawa coupling constants.
Expected precision of the $e$-$\mu$-$\varphi$ $(\varphi=A,H)$ measurements
at the $e$-gamma collider can be better than those at MEG and COMET
experiments. For $e$--$\tau$ mixing, its experimental reach is much
greater than those at LHCb and SuperKEKB. In addition,
chirality of the Higgs boson mediated  LFV coupling can be measured separately.
Therefore, the $e\gamma$ collider can be an useful tool to
measure the Higgs associated LFV.
Detailed phenomenological study will be elsewhere~\cite{Ref:future}.

\noindent
{\it Acknowledgments}~~~\\[2mm]
This work was supported, in part, by Japan Society for the Promotion
of Science, No.~18034004. 

\end{document}